\documentclass[aps,prl,twocolumn,amsmath,amssymb,amsfont,superscriptaddress,nofootinbib]{revtex4-1}
\usepackage{amsmath,aas_macros,xcolor,graphicx}
\newcommand{\bs}{\boldsymbol}
\newcommand{\diff}{{\mathrm d}}

\newcommand{\T}{\bold{T}}
\newcommand{\I}{\bold{I}}
\newcommand{\spin}{\bs{j}}
\newcommand{\Msun}{M_\odot}

\definecolor{darkgreen}{RGB}{50,150,0}

\DeclareMathAlphabet\mathbfcal{OMS}{cmsy}{b}{n}

\begin{document}
\title{Probing Primordial Chirality with Galaxy Spins}

\author{Hao-Ran~Yu}\email{\url{haoran@xmu.edu.cn}}
\affiliation{Department of Astronomy, Xiamen University, Xiamen, Fujian 361005, China}
\affiliation{Canadian Institute for Theoretical Astrophysics, 
University of Toronto, M5S 3H8, ON, Canada}
\affiliation{Tsung-Dao Lee Institute, Shanghai Jiao Tong University, Shanghai, 200240, China}

\author{Pavel Motloch}
\affiliation{Canadian Institute for Theoretical Astrophysics, 
University of Toronto, M5S 3H8, ON, Canada}

\author{Ue-Li~Pen}
\affiliation{Canadian Institute for Theoretical Astrophysics, 
University of Toronto, M5S 3H8, ON, Canada}
\affiliation{Tsung-Dao Lee Institute, Shanghai Jiao Tong University, Shanghai, 200240, China}
\affiliation{Dunlap Institute for Astronomy and Astrophysics, 
University of Toronto, M5S 3H4, ON, Canada}
\affiliation{Canadian Institute for Advanced Research, 
CIFAR Program in Gravitation and Cosmology, Toronto, M5G 1Z8, ON, Canada}
\affiliation{Perimeter Institute for Theoretical Physics, Waterloo, N2L 2Y5, ON, Canada}

\author{Yu~Yu}
\affiliation{Department of Astronomy, Shanghai Jiao Tong University, Shanghai, 200240, China}

\author{Huiyuan Wang}
\affiliation{Key Laboratory for Research in Galaxies and Cosmology, Department of Astronomy, 
University of Science and Technology of China, Hefei, Anhui 230026, China}

\author{Houjun Mo}
\affiliation{Tsinghua Center of Astrophysics \& Department of Physics, 
Tsinghua University, Beijing, 100084, China}
\affiliation{Department of Astronomy, University of Massachusetts Amherst, MA 01003, USA}

\author{Xiaohu Yang}
\affiliation{Department of Astronomy, Shanghai Jiao Tong University, Shanghai, 200240, China}
\affiliation{Tsung-Dao Lee Institute, Shanghai Jiao Tong University, Shanghai, 200240, China}

\author{Yipeng Jing}
\affiliation{Department of Astronomy, Shanghai Jiao Tong University, Shanghai, 200240, China}
\affiliation{Tsung-Dao Lee Institute, Shanghai Jiao Tong University, Shanghai, 200240, China}


\begin{abstract}
Chiral symmetry is maximally violated in weak interactions \citep{1956PhRv..104..254L}, 
and such microscopic asymmetries in the early Universe 
might leave observable imprints on astrophysical scales 
without violating the cosmological principle.
In this Letter, we propose a helicity measurement to detect primordial chiral violation.
We point out that observations of halo-galaxy angular momentum directions (spins), 
which are frozen in during the galaxy formation process, provide a fossil chiral observable. 
From the clustering mode of large scale structure of the Universe, 
we construct a spin mode in Lagrangian space and show in simulations that it is
a good probe of halo-galaxy spins.
In standard model, a strong symmetric correlation between the left 
and right helical components of this spin mode and galaxy spins is expected.  
Measurements of these correlations will be sensitive to chiral breaking, 
providing a direct test of chiral symmetry breaking in the early Universe.

\end{abstract}


\maketitle

Nature was originally considered to be simple and symmetric. 
Inquisitive investigations lead to the discovery of parity ($P$) and charge ($C$) 
violations in electroweak interactions \citep{1956PhRv..104..254L}. 
As chiral asymmetry exists on microscopic scales, it is interesting to ask whether the
early Universe also exhibited a chiral asymmetry 
\citep{1999PhRvL..83.1506L,2008PhRvL.101n1101C,2009PhRvL.102w1301T,
2016PhRvD..93k3002D,2017PhRvL.118v1301M}.
If present, it might be manifested as a primordial helicity violation.

To measure certain asymmetries one needs corresponding degrees of freedom (d.o.f). 
At the linear order, the large scale structure of 
the Universe is driven by a growing mode of the scalar perturbation, 
which corresponds to the convergence of the velocity field
and thus the growth of the density field.
This {\it clustering mode} ($E$ mode) has a single d.o.f
and cannot carry any primordial chiral-helicity violations. 
Beyond the linear order,
we may consider a primordial 3D vector field or a vector mode 
which can be decomposed into a scalar mode and two nonvanishing 
left-handed and right-handed helicity modes \citep{supplement_1}.
However, the observation is very challenging for such a primordial vector mode 
because, according to the linear perturbation theory, it decays away due to the
expansion of the Universe \citep{2002PhR...367....1B}.
If we hope to observe a primordial chiral imprints at low redshifts, 
a nondecaying vector mode should be reconstructed
to carry possibly frozen-in primordial chiral asymmetries.

Here, we construct such a primordial {\it spin mode}, a vector field
written as a quadratic function of the initial tidal field that is driven by the 
interaction between two linear {\it clustering modes} on two different scales; 
the persistent $E$ mode clustering enables this spin mode growing linearly with time.
This primordial spin mode well describes the vector field of angular momenta 
of protohalos and protogalaxies in Lagrangian space.
The $E$ mode clustering then maps this field to Eulerian space.
It is then possible to make a direct measurement of the primordial spin mode via
observations of the rotation directions of galaxies.
The spin mode has nontrivial curl and 
finding any asymmetry between properties of its left-handed and right-handed modes
would represent a detection of primordial chiral violation.

\textit{Clustering and spin modes. ---}
On large scales, the matter displacement field is close to being curl free and
the related single d.o.f, called clustering mode or $E$ mode, well captures the
inflow of matter into gravitational potential wells and ensuing increase in the density
contrast \citep{2014PhRvD..89h3515C,2017PhRvD..95d3501Y}.
Many techniques can reliably estimate this $E$ mode from the late stage large scale structure and 
thus reconstruct the cosmic initial conditions 
\citep{2014ApJ...794...94W,2017PhRvD..96l3502Z,2017MNRAS.469.1968P,2017ApJ...847..110Y,2017ApJ...841L..29W,2018PhRvD..97d3502Z,2019ApJ...870..116W}. 
The interaction between tidal fields on different scales can cause deviation from a pure
$E$ mode displacement and generate a divergence free $B$ mode.

In tidal torque theory \citep{1984ApJ...286...38W}, 
the initial angular momentum vector of a protohalo
that initially occupies Lagrangian volume $V_L$ is approximated by 
$j_\alpha\propto\epsilon_{\alpha \beta \gamma}I_{\beta \kappa}T_{\kappa\gamma}$, 
where $\I=(I_{\beta\kappa})$ is the moment of inertia tensor of $V_L$, 
$\T=(T_{\kappa\gamma})$ is the tidal tensor acting on $\I$, 
and $\epsilon_{\alpha \beta \gamma}$ is the Levi-Civita symbol collecting 
the antisymmetric components generated by the misalignment between $\I$ and $\T$. 
At late times, protohalos collapse into virialized systems and their angular momentum
decouples from the expansion of the Universe.
The spin directions are then frozen in these systems due to the angular momentum conservation.
Since $V_L$ represents a collection of matter that eventually clusters into 
a virialized system (dark matter halo), 
its initial collapse is driven by $\T$. As a result, $\I$ is closely aligned with $\T$ 
\citep{2000ApJ...532L...5L,2001ApJ...555..106L,2002MNRAS.332..325P}. 
If they were perfectly aligned, no spin would be generated in the leading order.
However,
some misalignment arises from inhomogeneity of the tidal field.
This motivates us to define
\begin{equation}\label{eq.spin1}
	\spin_R=(j_\alpha)\propto\epsilon_{\alpha \beta \gamma}\mathcal{T}_{\beta \kappa}
	\mathcal{T}^+_{\kappa\gamma},
\end{equation}
where $\mathbfcal{T}, \mathbfcal{T}^+$ are tidal fields constructed as Hessians 
of the initial gravitational potential smoothed at two different scales $r, R$. 
In what follows, we show that for properly chosen $r$ and $R$, 
$\spin_R$ is a good approximation for an angular momentum of a protohalo.
To obtain $\mathbfcal{T}, \mathbfcal{T}^+$, 
we smooth the initial gravitational potential $\phi_{\rm init}({\bs q})$, 
or its value obtained by reconstruction,
by multiplying it in the Fourier space by the baryonic acoustic oscillation damping
model ${\mathcal D}(k)^{1/4}=\exp(-r^2 k^2/2)^{1/4}$ \citep{2017PhRvD..95d3501Y}.

\textit{Helicity decomposition. ---}
Chirality can be violated without violating the cosmological principle (homogeneity and isotropy). 
In a simple analogy, sea snails could be distributed homogeneously in the sea
and each of them could be oriented in a random direction, 
however more than 90\% of their shells have right-handed whorls
\citep{Schilthuizen2005TheCE}.
The cosmic velocity and angular momentum fields could also be homogeneous and
isotropic on large scales but show local scale-dependent helicities. 
To quantify this, a {\it scale-dependent helicity decomposition} is required.
In Fourier space and for each nonzero $\bs{k}$ mode,
any vector field can be decomposed into a curl free and a pure 
left-helical and pure right-helical components \citep{supplement_1}. 
Angle-averaged, component-wise power spectra from these $k\neq 0$ Fourier modes
capture the statistics of helical asymmetries while satisfying the cosmological
principle, like the angle-averaged matter power spectrum $P(k)$.

The reconstructed spin mode $\spin_R$ is divergence free, 
and can then be decomposed into the left and right helical fields,
\begin{equation}\label{eq.helicity}
\spin_R=j_R^L \hat{\bs e}^L + j_R^R \hat{\bs e}^R; \ \ \ \ j_R^{L/R}\equiv\hat{\bs e}^{L/R}\cdot\spin_R,
\end{equation}
where the helical basis vectors $\hat{\bs e}^{L,R}$ are the eigenvectors 
of the curl operator.
In a helically symmetric Universe, the two helicities are statistically interchangeable and
the observed galaxy spins should have equal correlations to both.
A primordial parity violation (e.g., initial velocity with nonzero helicity) 
would affect $\spin_R^{L/R}$ unequally, which would lead to potentially observable effects.

\textit{Galaxy spins. ---}
The angular momentum vector of galaxies $\spin$ contains the magnitude $|\spin|$ 
and direction $\hat\spin$. While the magnitudes have been studied extensively, 
they are hard to measure and not predictable. We thus consider only directions 
(hereafter spin) of galaxy angular momenta as a viable tracer of primordial spin mode.
These parity-odd spins are observable (see discussions in 
\cite[Sec.II]{2019PhRvD..99l3532Y}), and the galaxy spin field is closely related to
the intrinsic alignment studies in that the minor axes of galaxies,
being a parity-even observable, are potentially 
probing the direction of the (unoriented) line given by $\{{\hat\spin},-\hat\spin\}$.

It is important to understand halo-galaxy spin connection.
It is conceivable that processes like cooling and feedback 
can decorrelate the spin directions of galaxies and their halos. 
However, in many recent hydrodynamic galaxy formation simulations 
(\citep{2015ApJ...812...29T,2019PhRvD..99l3532Y} and
references therein), which include various baryonic effects, we see a strong
halo-galaxy spin correlation. 
Quantitatively, the spin correlation is expressed by $\mu=\cos\theta$,
where $\theta\in[0^\circ,180^\circ]$ is the misalignment angle between halo and galaxy spins.
In these galaxy formation simulations, a typical $\langle\mu\rangle$ is 0.7,
much stronger than the null correlation $\langle\mu\rangle=0$.
As we do not see the halo-galaxy spin correlation erased by the baryonic effects,
we focus only on spins of dark matter halos in what follows.

\begin{figure}[t]
\centering
  \includegraphics[width=1\linewidth]{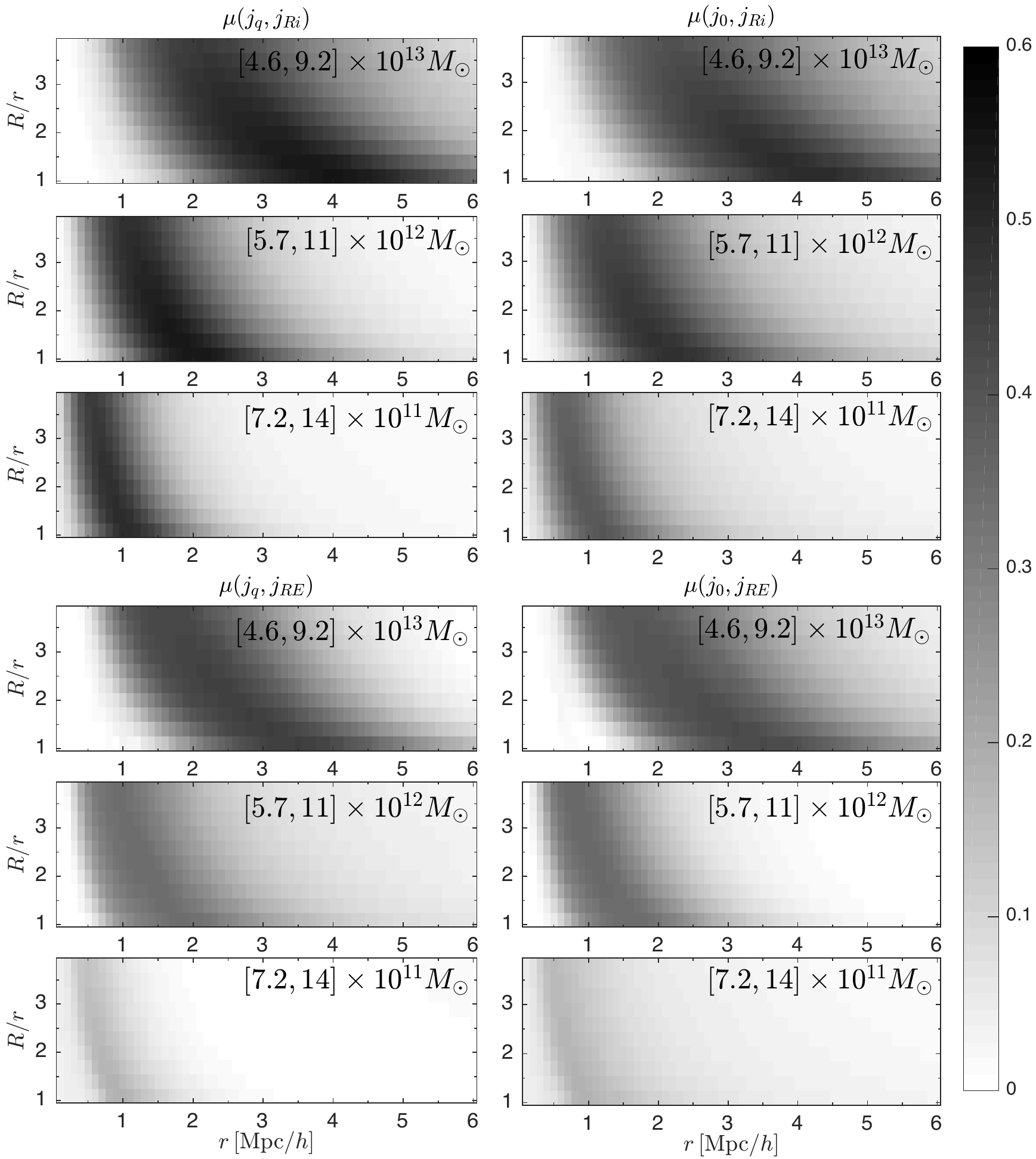}
\caption{
The cross-correlation coefficient between the Lagrangian $\spin_q$ (left) / Eulerian
$\spin_0$ (right) halo spins and spins reconstructed using Eq.\eqref{eq.spin1} from
known initial conditions
$\spin_{Ri}$ (top); $E$ mode reconstruction $\spin_{RE}$ (bottom). Plotted as a function of
the smoothing scales $r, R$, for three halo mass bins. Darker colors show better reconstruction of halo spin. 
 }\label{fig.1}
\end{figure}

\textit{Simulations and spin correlations.---}
We use numerical simulations to model a standard helically symmetric Universe 
and study the spin reconstruction.
We study two $N$-body simulations using the code {\small CUBE} \citep{2018ApJS..237...24Y},
that implements a particle-particle particle-mesh (P3M) algorithm.
One simulation is characterized by a periodic box size $L=200\, {\rm Mpc}/h$, 
grid number $N_g=768$ per dimension and a total number of dark matter particles $N_p=N_g^3$, 
and the other by $L=100\, {\rm Mpc}/h$, $N_g=400$ and $N_p=N_g^3$. 
Both simulations assume a WMAP5 cosmology \citep{2009ApJS..180..306D} and use Zel'dovich approximation 
\citep{1970A&A.....5...84Z} to generate initial conditions at redshift $z_i=50$. 
Their mass resolutions are close to $M_{\rm particle}\simeq 1.8\times 10^9\,\Msun$ 
and $M_{\rm particle}\simeq 1.6\times 10^9\,\Msun$. 
We identify dark matter halos with at least 100 particles 
($M_{\rm halo}\gtrsim 1.8\times 10^{11}\,\Msun$ and $M_{\rm halo}\gtrsim 1.6\times 10^{11}\,\Msun$) 
by a spherical overdensity (SO) algorithm. 
We compare their spins in the Lagrangian space,
	$\spin_q\propto\int_{V_L}(\bs{q}-\langle\bs{q}\rangle)\times(-\bs{\nabla}_q\phi)\diff^3\bs{q}$,
and in the Eulerian space (at redshift $z=0$),
	$\spin_0\propto\int_{V_E}\rho(\bs{x})(\bs{x}-\langle\bs{x}\rangle)\times\bs{v}(\bs{x})\diff^3\bs{x}$,
with the reconstructed spin.

We cross-correlate the halo spins $\spin_q$ and $\spin_0$ 
with the reconstructed spin field at the Lagrangian center-of-mass of these halos $\spin_R(\bs{q}_h)$. 
We quantify the cross-correlation of two vectors $\spin_A$ and $\spin_B$ by the cosine 
of their misalignment angle, $\mu(\spin_A,\spin_B)\equiv \spin_A\cdot\spin_B/|\spin_A||\spin_B|$,
averaged over halos in the given mass bin. 
In Fig.\ref{fig.1} we show cross-correlation coefficients between four combinations of
spins for a range of smoothing scales $r, R$ and halos in three mass bins.
The three subpanels in the top-left quarter of Fig.\ref{fig.1} show $\mu(\spin_q,\spin_{Ri})$, 
i.e., the cross-correlation between the Lagrangian spins of halos $\spin_q$ 
and the spin field $\spin_{Ri}$ reconstructed from the known initial conditions.
We find ``sweet spots'' in parameter space $(r,R)$ that optimize the cross-correlation. 
In general, the sweet spots lay in stripes and we find that 
choosing $R$ close to $r$ gives the strongest correlation. 
As expected, for more massive halos the sweet spots shift to larger $r$. 
The top-right quarter of Fig.\ref{fig.1} shows the same, 
only $\spin_{Ri}$ is correlated with the Eulerian spin $\spin_0$. 
The bottom part of Fig.\ref{fig.1} then shows correlations with $\spin_{RE}$, constructed using
Eq.\eqref{eq.spin1} from the initial potential as determined from the $E$ mode clustering,
$\phi_E\equiv \nabla_q^{-2}(-\nabla_q\cdot{\bs \Psi}({\bs q}))$.
As described in \cite{2017PhRvD..95d3501Y}, 
the $E$ mode clustering reconstruction is based on the divergence of the true displacement field ${\bs \Psi}$, 
and represents an upper limit of how well we can reconstruct the initial conditions of the
Universe when neglecting the $B$ mode portion of the displacement. 

From Fig.\ref{fig.1} we conclude that, 
(1) the best spin reconstruction is achieved with $R\rightarrow r_+$, 
(2) for more massive halos, larger $r$ is preferred, 
(3)  using $\spin_0$ 
instead of $\spin_q$ decreases the correlations only negligibly 
(because $\spin_q$ is highly correlated with $\spin_0$)
and (4) using $\phi_E$ instead of the true initial potential degrades the spin reconstruction, 
especially for less massive halos.
\begin{figure}
\centering
  \includegraphics[width=1.0\linewidth]{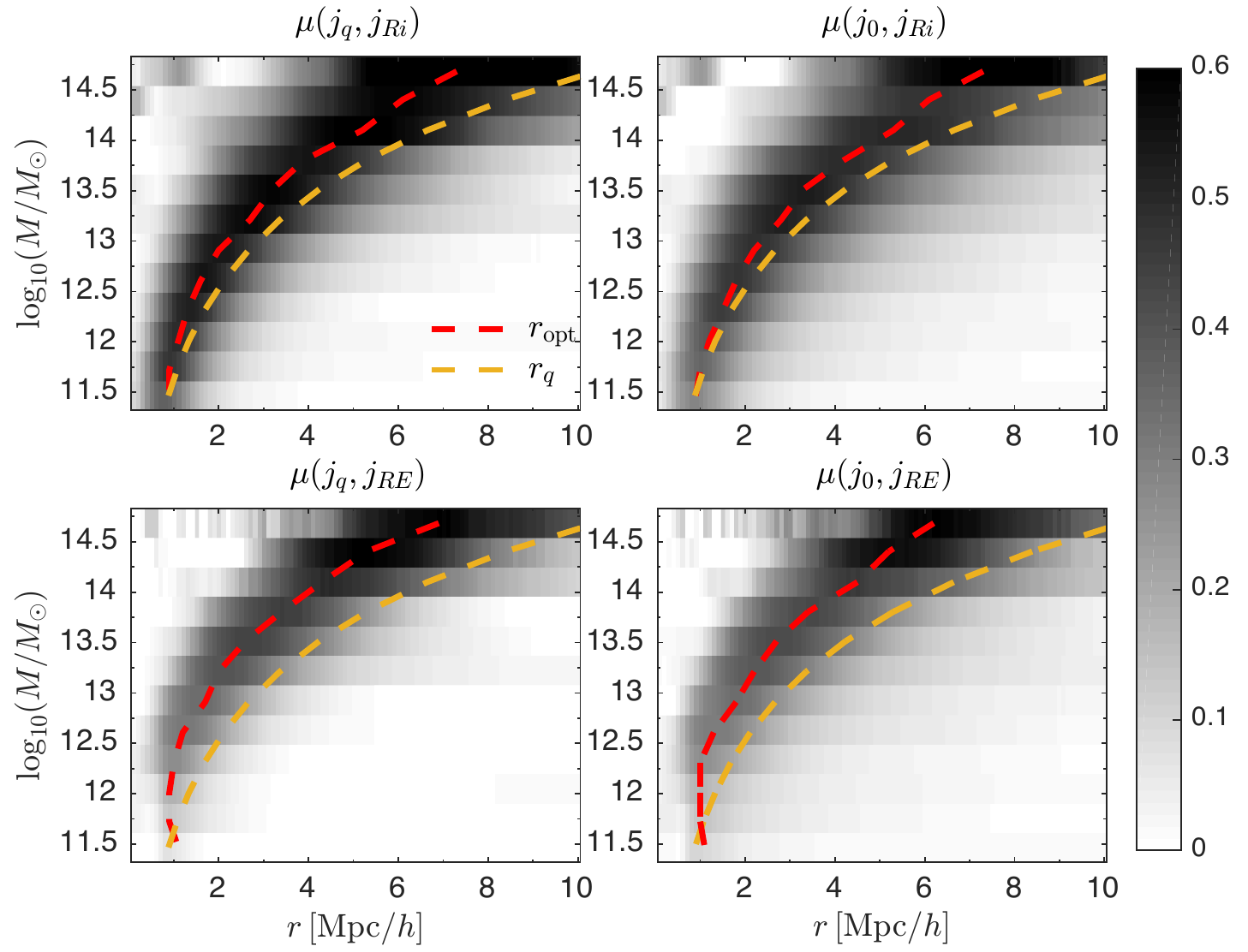}
 \caption{Optimal smoothing scale for spin reconstruction as a function of halo mass. 
 The four panels show the cases of cross-correlating $\spin_q$ (left) and $\spin_0$
 (right) with $\spin_{Ri}$ (top) and $\spin_{RE}$ (bottom). 
 Darker colors show better reconstruction. 
 Also plotted are the optimal smoothing scale $r_{\rm opt}$ (red dashed) and the
 Lagrangian radius $r_q$ of the halo (yellow dashed).}\label{fig.2}
\end{figure}

Based on the above investigations, we summarize the optimal smoothing scale as a function of halo mass in Fig.\ref{fig.2}. 
We chose $R\rightarrow r_+$ and plot $\mu$ as a function of halo mass $M$ and smoothing scale $r$
for each combination of spins
from Fig.\ref{fig.1}. The red dashed curves represent the optimal smoothing scale $r_{\rm
opt}$
and clearly show that $r_{\rm opt}$ increases with halo mass.
For reference, we also plot the equivalent protohalo radius in the Lagrangian space 
$r_q\equiv (2MG/\Omega_m H_0^2)^{1/3}$; in all cases
$r_{\rm opt}$ is close to $r_q$ but somewhat smaller. 
At a given halo mass, the optimal smoothing scale $r_{\rm opt}$ is smaller for
reconstruction using the $E$ mode clustering ($\spin_{RE}$) than for the one using the true
initial conditions ($\spin_{Ri}$).
This is probably caused by the
intrinsic smoothing effect in the $E$ mode reconstruction \citep{2017PhRvD..95d3501Y}. Furthermore, for
$\spin_{RE}$ the optimal $r_{\rm opt}$ is bounded by $r_{\rm opt}\gtrsim 1\, {\rm Mpc}/h$ 
even for low halo masses, 
which indicates loss of information below this scale.

\begin{figure}
\centering
  \includegraphics[width=0.8\linewidth]{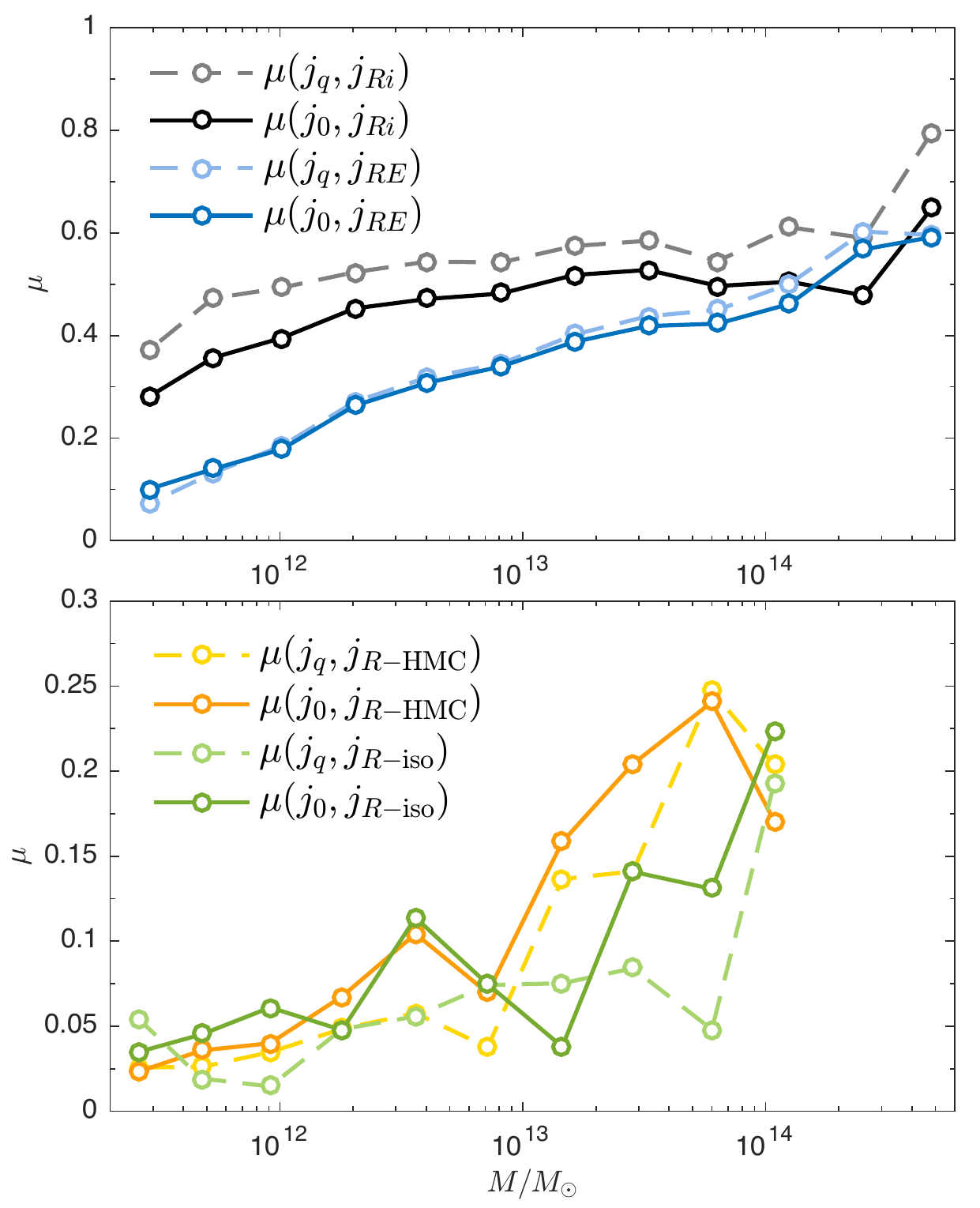}
 \caption{Accuracy of the spin reconstruction as a function of halo mass. 
 The Lagrangian spin $\spin_q$ and Eulerian spin $\spin_0$ are cross-correlated 
 with the spin reconstructions from, respectively, initial conditions $\spin_{Ri}$, 
 $E$ mode $\spin_{RE}$ (upper panel), HMC reconstruction $\spin_{R{\rm -HMC}}$ 
 and isobaric reconstruction $\spin_{R{\rm -iso}}$ (lower panel).}\label{fig.3}
\end{figure}

In the upper panel of Fig.\ref{fig.3}, we summarize the maximal achievable cross-correlation $\mu$
 as a function of the halo mass $M$.
As expected, using the true initial conditions to reconstruct the initial spin of protohalos gives the best result. 
Because reconstruction based on Eq.\eqref{eq.spin1} well captures the nature of spinning modes
in the Lagrangian space, the correlation is high:
$\mu(\spin_q,\spin_{Ri})$ is close to 0.5 at halo mass $M\simeq 10^{12}\Msun$. 
Since for halos in mass bins above $8 \times 10^{11} \Msun$ we observe average $\mu(\spin_q,\spin_0)\gtrsim 0.6$, 
the spins of halos are well conserved during evolution and
$\mu(\spin_0,\spin_{Ri})$ is comparably high. 
Using the $E$ mode decreases the quality of spin reconstruction for less massive halos, 
for which optimal reconstruction is achieved with smaller $r_{\rm opt}$. Their spins are thus sensitive to smaller
scale information, which is gradually lost in the $E$ mode reconstruction.
As a result, $\mu(\spin_0,\spin_{RE})$ drops below 0.2 at $M\simeq 10^{12}\Msun$. 

To get a better estimate of what is realistically possible, we use the small box
simulation to reconstruct the initial gravitational potential using either Hamiltonian Markov chain (HMC) 
reconstruction \citep{2014ApJ...794...94W} 
or the isobaric reconstruction \citep{2017PhRvD..96l3502Z,2017MNRAS.469.1968P}.
These gravitational potentials are then used to reconstruct the initial spin field according to Eq.\eqref{eq.spin1}. 
We plot the results of spin reconstruction from HMC and isobaric density reconstructions 
in the lower panel of Fig.\ref{fig.3}. 
Because they lose even more small scale information, the spin correlations are around 0.05
$-$ 0.1 
for halo masses between $10^{12}\Msun$ and $10^{13}\Msun$. 
For both isobaric and HMC results, we find that $r_{\rm opt}\simeq 2.5 \, {\rm Mpc}/h$, 
which indicates that the potential fields $\phi_{\rm iso}$ and $\phi_{\rm HMC}$ contain little 
information below this scale and one has to use larger scale tidal field to reconstruct the spin.

\textit{Conclusion and discussion.---}
We considered the possibility of primordial chiral violation of the Universe.
To facilitate a detection,
we construct a nondecaying vector mode which is frozen in during the epoch of galaxy formation, 
and can be carried to the current epoch of the Universe by galaxy spins. 
Any primordial chiral asymmetries projected to the helical decomposition 
[Eq.\eqref{eq.helicity}] of the spin mode can be reflected in the observed galaxy spins.

Vector modes are a combination of left and right helical mode. 
They can arise primordially, through second order interactions 
in the standard model \citep{2009JCAP...02..023L}, 
or through interactions beyond the standard model.  
Cosmic strings, for example, are active, causal sources of vector modes \citep{1998PhRvD..58b3506T}.
The standard Nambu action for the equation of motion is parity symmetric, 
but one could envision non-Abelian helicity-violating equations of motion \citep{1997ApJ...491L..67S}.
On scales of $\sim 1\,{\rm Mpc}$ their amplitude is very poorly constrained, 
and they could annihilate after imprinting helicity on the velocity field.  
A modified version of gravity could include a source term $\boldsymbol{F}=H\epsilon\bold{T}\bold{T}^+$, 
where $H$ is an explicitly helicity violating operator. 
Such a term appears nonlocal, 
though no more than the nonlocality of Poisson’s equation for the Newtonian limit.  
We leave it for future work to consider the extension to a causal, 
relativistic theory with such a slow motion limit.

A measurement from survey data is beyond the scope of this Letter.
The observational complexities require a thorough understanding of the systematics.
The strong spin correlation between disk galaxies and their parent halos is confirmed 
in hydrodynamic simulations \citep{2015ApJ...812...29T,2019MNRAS.488.4801J}, 
so in this Letter we use halos to represent galaxies. 
Strong Lagrangian-Eulerian spin correlation has also been confirmed in 
this work and many high-resolution $N$-body simulations
\citep{2000ApJ...532L...5L,2001ApJ...555..106L,2002MNRAS.332..325P,2019PhRvD..99l3532Y}. 
Both of these correlations ($\left\langle\mu\right\rangle>0.6$) 
are reliably above the null correlation $\left\langle\mu\right\rangle=0$,
indicating that neither the nonlinear effects at low redshift, 
nor baryonic effects during galaxy formation has washed out the primordial spin modes. 
For the halo-galaxy spin direction correlation $\langle\mu\rangle=0.6$,
the number of galaxies needed for a spin-correlation detection increases by 
only $\langle\mu\rangle^{-1}-1\simeq 67\%$, compared to the case that galaxy spins perfectly
trace the spin direction of their halos ($\mu=1$).

The $E$ mode reconstruction, including complications from the bias of tracers 
\citep{2017ApJ...847..110Y,2017ApJ...841L..29W}, redshift space distortions and survey masks 
\citep{2018PhRvD..97d3502Z} has been intensively discussed in literature. 
{\small ELUCID} \citep{2016ApJ...831..164W}, 
state-of-the-art constrained simulation of the real Universe 
with all the above effects considered, can in principle be used for spin reconstruction. 
To compare with galaxy rotations, 
one needs an inverse displacement mapping from redshift space to Lagrangian space, 
with the uncertainties arising from redshift space distortion reconstruction and shell crossing, among other
effects. 
The error in the estimation of parent halo mass is about a factor of 2
\citep{1975BAAS....7..426T}, comparable to the width of the halo 
mass bins used in this Letter.
We find that such uncertainty in halo mass is sufficient to determine
the optimal smoothing scale $r_{\rm opt}$ with good accuracy.

From simulations we have estimated that the correlation between halo spins and the corresponding prediction 
from estimators quadratic in the reconstructed displacement fields should be over 10\%.  
Large samples of galaxy spins are thus expected to result in a significant detection of correlation.
This opens the opportunity to test helicity violation observationally.

Two tests are amenable to observations: 
(1) the galaxy spin field can be decomposed into left and right helical fields, and their
respective correlations with the reconstructed spin mutually compared;  
(2) the galaxy spin can be correlated with the left and right helical projections of the
reconstructed spin. A signal in either scenario implies a primordial helical vorticity. 

For a galaxy of negligible thickness, direction of its angular momentum can be deduced
from the observed position angle and axis ratio. This limit is well applicable to spiral
galaxies; correction for finite thickness can be applied \citep{1984AJ.....89..758H}. The
citizen science project Galaxy Zoo \citep{2008MNRAS.388.1686L} classified tens of
thousands of spiral galaxies (see also \citep{2011PhLB..699..224L} for an independent
classification effort), which can be thus utilized to search for primordial helicity
violation using the ideas presented in this work. A statistically significant preference
for $S$-wise winding galaxies in this sample was found to be a selection effect
\citep{2017MNRAS.466.3928H}. 
This artificial signal contributes to the $k=0$ mode of the helicity decomposition 
and violates the cosmological principle. 
It might increase noise in the measurement but is not expected to bias the results.

Helicity is a quadratic function of spin in Fourier space \citep{supplement_1}, 
so the chiral asymmetry of field $-\spin$ is the same as that of $\spin$. 
Intrinsic alignment of galaxies are closely related to the direction of the (unoriented) 
line given by $\{{\hat\spin},-\hat\spin\}$ and potentially also have primordial 
chiral violation frozen in. We leave it to future studies.

\textit{Acknowledgments.---}
H.R.Y., P.M. and U.L.P. acknowledge the funding from NSERC.
H.R.Y. is supported by NSFC 11903021. H.Y.W. is supported by NSFC 11421303.
The simulations were performed on the supercomputer at CITA.

\bibliography{haoran_ref}

\begin{thebibliography}{10}

\bibitem{1956PhRv..104..254L}
T.~D. {Lee} and C.~N. {Yang},
\newblock Physical Review {\bf 104}, 254 (1956).

\bibitem{1999PhRvL..83.1506L}
A.~{Lue}, L.~{Wang}, and M.~{Kamionkowski},
\newblock \prl {\bf 83}, 1506 (1999), astro-ph/9812088.

\bibitem{2008PhRvL.101n1101C}
C.~R. {Contaldi}, J.~{Magueijo}, and L.~{Smolin},
\newblock \prl {\bf 101}, 141101 (2008), 0806.3082.

\bibitem{2009PhRvL.102w1301T}
T.~{Takahashi} and J.~{Soda},
\newblock \prl {\bf 102}, 231301 (2009), 0904.0554.

\bibitem{2016PhRvD..93k3002D}
G.~{Dvali} and L.~{Funcke},
\newblock \prd {\bf 93}, 113002 (2016).

\bibitem{2017PhRvL.118v1301M}
K.~W. {Masui}, U.-L. {Pen}, and N.~{Turok},
\newblock \prl {\bf 118}, 221301 (2017), 1702.06552.

\bibitem{supplement_1}
See the {\it Supplemental Material} of this Letter for the scale-dependent
  helicity decomposition .

\bibitem{2002PhR...367....1B}
F.~{Bernardeau}, S.~{Colombi}, E.~{Gazta{\~n}aga}, and R.~{Scoccimarro},
\newblock \physrep {\bf 367}, 1 (2002), astro-ph/0112551.

\bibitem{2014PhRvD..89h3515C}
K.~C. {Chan},
\newblock \prd {\bf 89}, 083515 (2014), 1309.2243.

\bibitem{2017PhRvD..95d3501Y}
H.-R. {Yu}, U.-L. {Pen}, and H.-M. {Zhu},
\newblock \prd {\bf 95}, 043501 (2017), 1610.07112.

\bibitem{2014ApJ...794...94W}
H.~{Wang}, H.~J. {Mo}, X.~{Yang}, Y.~P. {Jing}, and W.~P. {Lin},
\newblock \apj {\bf 794}, 94 (2014), 1407.3451.

\bibitem{2017PhRvD..96l3502Z}
H.-M. {Zhu}, Y.~{Yu}, U.-L. {Pen}, X.~{Chen}, and H.-R. {Yu},
\newblock \prd {\bf 96}, 123502 (2017).

\bibitem{2017MNRAS.469.1968P}
Q.~{Pan}, U.-L. {Pen}, D.~{Inman}, and H.-R. {Yu},
\newblock \mnras {\bf 469}, 1968 (2017), 1611.10013.

\bibitem{2017ApJ...847..110Y}
Y.~{Yu}, H.-M. {Zhu}, and U.-L. {Pen},
\newblock \apj {\bf 847}, 110 (2017).

\bibitem{2017ApJ...841L..29W}
X.~{Wang} {\em et~al.},
\newblock \apjl {\bf 841}, L29 (2017), 1703.09742.

\bibitem{2018PhRvD..97d3502Z}
H.-M. {Zhu}, Y.~{Yu}, and U.-L. {Pen},
\newblock \prd {\bf 97}, 043502 (2018), 1711.03218.

\bibitem{2019ApJ...870..116W}
X.~{Wang} and U.-L. {Pen},
\newblock \apj {\bf 870}, 116 (2019), 1807.06381.

\bibitem{1984ApJ...286...38W}
S.~D.~M. {White},
\newblock \apj {\bf 286}, 38 (1984).

\bibitem{2000ApJ...532L...5L}
J.~{Lee} and U.-L. {Pen},
\newblock \apj {\bf 532}, L5 (2000), astro-ph/9911328.

\bibitem{2001ApJ...555..106L}
J.~{Lee} and U.-L. {Pen},
\newblock \apj {\bf 555}, 106 (2001), astro-ph/0008135.

\bibitem{2002MNRAS.332..325P}
C.~{Porciani}, A.~{Dekel}, and Y.~{Hoffman},
\newblock \mnras {\bf 332}, 325 (2002), astro-ph/0105123.

\bibitem{Schilthuizen2005TheCE}
M.~Schilthuizen and A.~Davison,
\newblock Naturwissenschaften {\bf 92}, 504 (2005).

\bibitem{2019PhRvD..99l3532Y}
H.-R. {Yu}, U.-L. {Pen}, and X.~{Wang},
\newblock \prd {\bf 99}, 123532 (2019), 1810.11784.

\bibitem{2015ApJ...812...29T}
A.~F. {Teklu} {\em et~al.},
\newblock \apj {\bf 812}, 29 (2015), 1503.03501.

\bibitem{2018ApJS..237...24Y}
H.-R. {Yu}, U.-L. {Pen}, and X.~{Wang},
\newblock The Astrophysical Journal Supplement Series {\bf 237}, 24 (2018).

\bibitem{2009ApJS..180..306D}
J.~{Dunkley} {\em et~al.},
\newblock \apjs {\bf 180}, 306 (2009), 0803.0586.

\bibitem{1970A&A.....5...84Z}
Y.~B. {Zel'dovich},
\newblock \aap {\bf 5}, 84 (1970).

\bibitem{2009JCAP...02..023L}
T.~H.-C. {Lu}, K.~{Ananda}, C.~{Clarkson}, and R.~{Maartens},
\newblock \jcap {\bf 2009}, 023 (2009), 0812.1349.

\bibitem{1998PhRvD..58b3506T}
N.~{Turok}, U.-L. {Pen}, and U.~{Seljak},
\newblock \prd {\bf 58}, 023506 (1998), astro-ph/9706250.

\bibitem{1997ApJ...491L..67S}
D.~{Spergel} and U.-L. {Pen},
\newblock \apjl {\bf 491}, L67 (1997), astro-ph/9611198.

\bibitem{2019MNRAS.488.4801J}
F.~{Jiang} {\em et~al.},
\newblock \mnras {\bf 488}, 4801 (2019), 1804.07306.

\bibitem{2016ApJ...831..164W}
H.~{Wang} {\em et~al.},
\newblock \apj {\bf 831}, 164 (2016), 1608.01763.

\bibitem{1975BAAS....7..426T}
R.~B. {Tully}, O.~{de Marseille}, and J.~R. {Fisher},
\newblock {A New Method of Determining Distances to Galaxies},
\newblock in {\em Bulletin of the American Astronomical Society}, volume~7 of
  {\em \baas}, p. 426, 1975.

\bibitem{1984AJ.....89..758H}
M.~P. {Haynes} and R.~{Giovanelli},
\newblock \aj {\bf 89}, 758 (1984).

\bibitem{2008MNRAS.388.1686L}
K.~{Land} {\em et~al.},
\newblock \mnras {\bf 388}, 1686 (2008), 0803.3247.

\bibitem{2011PhLB..699..224L}
M.~J. {Longo},
\newblock Physics Letters B {\bf 699}, 224 (2011), 1104.2815.

\bibitem{2017MNRAS.466.3928H}
W.~B. {Hayes}, D.~{Davis}, and P.~{Silva},
\newblock \mnras {\bf 466}, 3928 (2017), 1610.07060.

\end{thebibliography}

\newpage

\appendix

\section{Supplemental Material}

In this supplement, we give the details of our helicity decomposition
and measurement calculation. A smooth 3D vector field can be decomposed into three helical components,
based on eigenvalues of the curl ($\nabla\times$) operator. In Fourier space, the
corresponding eigenequation reads
$i\epsilon_{\alpha\beta\gamma}k_{\beta}\tilde{u}_{\gamma}=\lambda\tilde{u}_\alpha$.  
For $k\equiv|\bs{k}|\neq 0$, the three eigenvalues read $\lambda^E=0$ and $\lambda^{L/R}=\mp
k$.  Here $\lambda^E$ defines an $E$-mode, non-helical, irrotational field, whereas
$\lambda^L$ and $\lambda^R$ correspond to pure left-helical and pure right-helical
fields. The three corresponding eigenbases $\hat{\bs{e}}^s(\bs{k})$ ($s=E,L,R$) are
orthogonal and complete.

The helical decomposition can be achieved by projecting the Fourier modes
$\tilde{\bs{u}}(\bs{k})$ onto $\hat{\bs{e}}^s(\bs{k})$; the corresponding projection
matrices ${\bold P}^s=\hat{\bs{e}}^s\hat{\bs{e}}^{s\dagger}$ are explicitly
\begin{eqnarray}
	{\bold P}^E_{\alpha\beta}&=&k_\alpha k_\beta/k^2,\\
	{\bold P}^{L/R}_{\alpha\beta}&=&\frac{1}{2}\left[(\delta_{\alpha\beta}-k_\alpha k_\beta/k^2)\pm i\epsilon_{\alpha\beta\gamma}k_\gamma/k\right].
\end{eqnarray}
Using them, a wave packet can be decomposed as
\begin{eqnarray}\label{eq.helicity_decomposition}
	&&(2\pi)^{-3}\tilde{\bs{u}}(\bs{k})e^{i\bs{k}\cdot\bs{x}}=
	(2\pi)^{-3}e^{i\bs{k}\cdot\bs{x}}\sum_s {\bold P}^s\tilde{\bs{u}}(\bs{k}) \nonumber \\
	&=&(2\pi)^{-3}e^{i\bs{k}\cdot\bs{x}}\sum_s\tilde{\bs{u}}^s(\bs{k})
	=\sum_s{\bs{u}}^s(\bs{x};\bs{k}),
\end{eqnarray}
where we have defined $\bs{u}^s(\bs{x};\bs{k})=(2\pi)^{-3}\tilde{\bs{u}}_s(\bs{k})e^{i\bs{k}\cdot\bs{x}}$. 

Helicity decomposition of an arbitrary vector field can be then accomplished through
rewriting it as a linear combination of wave packets, which can be then individually
decomposed as in \eqref{eq.helicity_decomposition}.
As a result, a general vector field $\bs{u}(\bs{x})$ can be written as
	$\bs{u}=\bs{u}^E+\bs{u}^L+\bs{u}^R$,
sum of an $E$-mode (including the constant $k=0$ mode), $L$-mode and an $R$-mode field.

The helicity of a 3D vector field $\bs{u}$ inside volume $V$ is defined as 
$H=\int \mathcal{H}\bs{u}\diff V=\int\bs{u}\cdot(\nabla\times\bs{u})\diff V$
and captures the mutual orientation between a field and its vorticity.
We also define $H(k)\equiv\int\mathcal{H}\bs{u}(k)\diff V$ the 
{\it scale-dependent helicity} of $\bs{u}$ on scale $k$, where 
$\bs{u}(k)=\sum_{\bs{k}=|k|}\tilde{\bs{u}}(\bs{k})e^{i\bs{k}\cdot\bs{x}}$ 
is the collection of wave packets of wavenumber $k$.
By using Parseval's theorem,
\begin{eqnarray}
	&&\int_V\mathcal{H}\bs{u}^{L/R}\diff^3x=\int_V\bs{u}^{L/R}\cdot\left(\nabla\times\bs{u}^{L/R}\right)\diff^3x \nonumber\\
	&=&\int\tilde{\bs{u}}^{L/R*}(\bs{k})\cdot\left(\mp k\tilde{\bs{u}}^{L/R}(\bs{k})\right)\diff^3k\\
	&=&\mp\int |\tilde{\bs{u}}^{L/R}(\bs{k})|^2 k\diff^3k \nonumber \\
	&=&\mp\int_0^\infty P^{L/R}(k)k\diff k,
\end{eqnarray}
where we defined $P^s(k)$ the power spectrum of the helicity component $\bs{u}^s$. 
Because $\int\mathcal{H}\bs{v}^E\diff^3x=0$ and the orthogonality of the helicity 
decomposition (Eq.(\ref{eq.helicity_decomposition})) we see $H(k)=k(P^R(k)-P^L(k))$.

The $\spin_R({\bs q})$ from Eq.(\ref{eq.spin1}) provides a spin mode that can be
determined by measuring the spin direction of galaxies/halos $\spin_{\rm obs}({\bs q})$.
As $\spin_R({\bs q})$ is divergence free\footnote{
$j_{R\alpha}\propto\epsilon_{\alpha\beta\gamma}
	\left(\partial_\beta\partial_\kappa\phi\right)
	\left(\partial_\kappa\partial_\gamma\phi^+\right)$,
where $\phi$ and $\phi^+$ are the gravitational potential smoothed on two scales to produce the tidal tensor field, and its divergence $\nabla\cdot\bs{j}_R=\partial_\alpha j_{R\alpha}
    \propto\partial_\alpha[\epsilon_{\alpha\beta\gamma}
	(\partial_\beta\partial_\kappa\phi)
	(\partial_\kappa\partial_\gamma\phi^+)]
	=
	\epsilon_{\alpha\beta\gamma}
	[(\partial_\alpha\partial_\beta\partial_\kappa\phi)
	(\partial_\kappa\partial_\gamma\phi^+)
	+
	(\partial_\beta\partial_\kappa\phi)
	(\partial_\alpha\partial_\kappa\partial_\gamma\phi^+)]=0$.},
it can be decomposed into a left and a right helical mode $\spin_R^{L/R}$.
The primordial asymmetry of $\spin_{\rm init}^{L/R}$ is projected onto $\spin_R^{L/R}$
and is eventually measured in the nonzero chiral asymmetry $H^{\rm obs}(k)$ of
$\spin_{\rm obs}^{L/R}$.  The evolution of parity violations can be schematically
summarized as
\begin{equation}
		\bs{j}_{\rm init}^{L/R}\xrightarrow[]{\rm projection}\bs{j}_R^{L/R}\xrightarrow[]{\rm evolution}\bs{j}_{\rm obs}^{L/R}.
\end{equation}
In numerical simulations we have detected that left and right helicity components of the
dark matter halo spins
arising from helicity violating initial conditions (represented by helicity violating
initial velocities $\bs{u}_{\rm init}$) have different properties, and we will present
these results in a future paper.

\end{document}